\documentclass[twocolumn,english]{revtex4-2}
\usepackage[T1]{fontenc}
\usepackage[latin9]{luainputenc}
\usepackage{babel}
\usepackage{float}
\usepackage{amsbsy}
\usepackage{amstext}
\usepackage{graphicx}
\usepackage[pdfusetitle,
 bookmarks=true,bookmarksnumbered=false,bookmarksopen=false,
 breaklinks=false,pdfborder={0 0 1},backref=false,colorlinks=false]
 {hyperref}
\begin{document}
\title{Persistent self-organized states in non-equilibrium magnetic models}
\author{R. A. Dumer }
\email{rafaeldumer@fisica.ufmt.br}

\affiliation{Programa de Pós-Graduação em Física, Instituto de Física, Universidade
Federal de Mato Grosso, Cuiabá, Brasil.}
\author{M. Godoy}
\email{mgodoy@fisica.ufmt.br}

\affiliation{Programa de Pós-Graduação em Física, Instituto de Física, Universidade
Federal de Mato Grosso, Cuiabá, Brasil.}
\begin{abstract}
In this work, we employed Monte Carlo simulations to study the Ising,
$XY$, and Heisenberg models on a simple cubic lattice, where the
system models evolve toward the steady state under the influence of
competition between one- and two-spin flip dynamics. With probability
$q$, the system is in contact with a thermal reservoir at temperature
$T$ and evolves toward the lower energy state through one-spin flip
dynamics. On the other hand, with probability $1-q$, the system is
subjected to an external energy flux that drives it toward the higher
energy state through two-spin flip dynamics. As a result, we constructed
the phase diagram of $T$ as a function of $q$. In this diagram,
we identified the antiferromagnetic ($AF$) ordered phase, the ferromagnetic
($F$) ordered phase, and the disordered paramagnetic ($P$) phase
for all the models studied. Through these phases, we observed self-organization
phenomena in the systems. For low values of $q$, the system is in
the $AF$ phase, and as $q$ increases the system continuously transitions
to the $P$ phase. Now, for high values of $q$, the system through
continuous phase transitions again reaches an ordered phase, the $F$
phase, at low values of $T$. Additionally, we also calculated the
critical exponents of the system, showing that these are not affected
by the non-equilibrium regime of the system.
\end{abstract}
\maketitle

\section{Introduction}

Magnetic models, such as Ising \citep{1}, Heisenberg \citep{2},
and $XY$ \citep{3}, play a central role in statistical mechanics
providing essential theoretical frameworks for studying phase transitions
and critical behavior. In the Ising model, spins are restricted to
a single axis, while in the $XY$ model, they are confined to a plane.
The Heisenberg model, in contrast, allows spins to adopt orientations
in three dimensions. Despite their versatility, continuous models
like $XY$ and Heisenberg exhibit limitations in low dimensions, as
established by the Mermin-Wagner theorem \citep{4}, which prohibits
magnetic ordering in two dimensions for short-range interactions at
finite temperatures. In higher dimensions, however, these constraints
vanish, and these models display characteristic phase transitions.

Self-organization is a hallmark of many non-equilibrium systems, where
macroscopic order emerges spontaneously from microscopic interactions
without external coordination \citep{5}. Examples include pattern
formation in chemical reactions, such as the Belousov-Zhabotinsky
reaction \citep{6}, biological systems like flocking in birds \citep{7},
cellular structures \citep{8}, and even traffic flow dynamics \citep{9}.
These systems demonstrate how competing processes and feedback mechanisms
can drive the system toward ordered states or complex behaviors. In
the context of magnetic systems, competition between different dynamics
can also lead to self-organization, fostering the emergence of non-trivial
steady states, which may exhibit transitions between different ordered
phases \citep{10,11,12,13}.

In dealing with magnetic systems governed by competing dynamics, the
principle of microscopic reversibility is not always satisfied, forcing
the system out of equilibrium and beyond the scope of the formalism
proposed by Gibbs for thermodynamic equilibrium systems \citep{14}.
In this competition, one-spin flip and two-spin flip dynamics are
frequently employed \citep{15,16,17,18}. The one-spin flip dynamic,
acting with probability $q$, simulates the system in contact with
a thermal reservoir at temperature $T$ and favors the lowest energy
state of the system. In contrast, the two-spin flip dynamic, acting
with probability $1-q$, introduces an external energy flux into the
system, favoring the highest energy state of the system.

In previous studies on the competition between these reactive dynamics
using the two-dimensional Ising model, both on regular \citep{15,16}
and complex \citep{17,18} networks, very similar phase diagrams were
found. For low values of $q$, an antiferromagnetic phase almost independent
of temperature is observed. As $q$ increases, a second-order phase
transition leads to the paramagnetic phase. However, with further
increases in $q$, for low temperature values, the system transitions
continuously back to an ordered phase, the ferromagnetic phase. This
process reflects the self-organization of the system.

However, no studies have yet explored the competition between these
dynamics in continuous-spin-state models. Thus, the objective of the
present work is to implement the competition between one- and two-spin
flip dynamics in the Ising, $XY$, and Heisenberg models on a simple
cubic lattice, keeping the lattice unchanged. By focusing on variations
in the magnetic models, we aim to verify whether the self-organization
phenomenon and the phase diagram topology observed in previous studies
persist under these new conditions.

This article is organized as follows: In Section \ref{sec:Model},
we provide details about the Monte Carlo method (MC), the thermodynamic
quantities of interest, and the scaling relations for each of them.
The results are discussed in Section \ref{sec:Results}, including
the phase diagram and description of the second-order phase transitions.
Finally, in Section \ref{sec:Conclusions}, we present the conclusions
drawn from the study.

\section{Model and Method\protect\label{sec:Model}}

The interaction energy between spins is defined by the Hamiltonian
model in the following form:

\begin{equation}
\mathcal{H}=-\sum_{\left\langle i,j\right\rangle }\left\{ J_{\perp}\left(S_{i}^{x}S_{j}^{x}+S_{i}^{y}S_{j}^{y}\right)+J_{\parallel}\left(S_{i}^{z}S_{j}^{z}\right)\right\} ,\label{eq:1}
\end{equation}
where the summation runs over all pairs of nearest neighbors on a
regular simple cubic lattice, and the spins are treated as three-component
vectors, with $\left|\boldsymbol{S}_{i}\right|=1$. In the special
case where $J_{\perp}=0$ and $J_{\parallel}=1$, the model reduces
to the Ising model, while for $J_{\parallel}=0$ and $J_{\perp}=1$,
it corresponds to the $XY$ model. For the most general case, where
$J_{\perp}=J_{\parallel}=1$, the system represents the 3D isotropic
Heisenberg model.

We have simulated the system specified by the Hamiltonian in Eq. (\ref{eq:1}),
employing MC simulations. We always assumed periodic boundary conditions.
Starting from the random initial state, a new spin configuration is
generated following the Markov process: for a given competition probability
$q$, temperature $T$ e system size $N$, we randomly select a site
on the lattice, $\boldsymbol{S}_{i}$, and generate a random number
$\xi$ between zero and one. If $\xi\le q$, one-spin flip dynamics
is chosen to simulate the system in contact with a heat bath. In this
case, we randomly choose a new state for $\boldsymbol{S}_{i}$, which
is accepted according to the Metropolis prescription \citep{19}: 

\begin{equation}
W(\boldsymbol{S}_{i}\to\boldsymbol{S}_{i}^{\prime})=\left\{ \begin{array}{cccc}
e^{\left(-\Delta E_{i}/k_{B}T\right)} & \textrm{if} & \Delta E_{i}>0\\
1 & \textrm{if} & \Delta E_{i}\le0 & ,
\end{array}\right.\label{eq:2}
\end{equation}
where $\Delta E_{i}$ is the change in energy, based in Eq. (\ref{eq:1}),
after flipping the spin $\boldsymbol{S}_{i}$, and $k_{B}$ is the
Boltzmann constant. With this transition rate, the new state is accepted
if $\Delta E_{i}\le0$, but if $\Delta E>0$ the state can still be
accepted if, upon generating a random number $\xi_{1}$ between zero
and one, it is less than or equal to the Boltzmann factor $\exp\left(-\Delta E_{i}/k_{B}T\right)$.
On the other hand, if $\xi>q$, two-spin flip dynamics is chosen to
simulate the system subjected to an external energy flux. In this
case, we also randomly select one of the nearest neighbors of $\boldsymbol{S}_{i}$,
$\boldsymbol{S}_{j}$. Here, both spins have their states altered
randomly, and these new states are accepted with the following rate:

\begin{equation}
W(\boldsymbol{S}_{i}\boldsymbol{S}_{j}\to\boldsymbol{S}_{i}^{\prime}\boldsymbol{S}_{j}^{\prime})=\left\{ \begin{array}{cccc}
0 & \textrm{if} & \Delta E_{ij}\le0\\
1 & \textrm{if} & \Delta E_{ij}>0 & ,
\end{array}\right.\label{eq:3}
\end{equation}
where $\Delta E_{ij}$ is the energy difference resulting from the
change in spin states. With this dynamic, we observe that the change
is only accepted if it increases the energy of the system, as expected
for a system subjected to an external energy flux.

Repeating the Markov process $N$ times constitutes one Monte Carlo
Step (MCS). We allowed the system to evolve for $10^{5}$ MCS to reach
a stationary state, for all lattice sizes. To calculate the thermal
averages of the quantities of interest, we conducted an additional
$9\times10^{5}$ MCS. The statistical errors were calculated using
the Bootstrap method \citep{20}. 

The measured thermodynamic quantities in our simulations are: magnetization
per spin $\textrm{m}_{\textrm{L}}^{\textrm{F}}$, staggered magnetization
$\textrm{\ensuremath{\textrm{m}_{\textrm{L}}^{\textrm{AF}}}}$, magnetic
susceptibility $\textrm{\ensuremath{\chi}}_{\textrm{L}}$ and reduced
fourth-order Binder cumulant $\textrm{U}_{\textrm{L}}$:

\begin{equation}
\textrm{m}_{\textrm{L}}^{\textrm{F}}=\sqrt{\left(m_{x}^{F}\right)^{2}+\left(m_{y}^{F}\right)^{2}+\left(m_{z}^{F}\right)^{2}},\label{eq:4}
\end{equation}

\begin{equation}
\textrm{m}_{\textrm{L}}^{\textrm{AF}}=\sqrt{\left(m_{x}^{AF}\right)^{2}+\left(m_{y}^{AF}\right)^{2}+\left(m_{z}^{AF}\right)^{2}},\label{eq:5}
\end{equation}

\begin{equation}
\textrm{\ensuremath{\chi}}_{\textrm{L}}=\frac{N}{k_{B}T}\left[\left\langle \textrm{m}_{\textrm{L}}^{2}\right\rangle -\left\langle \textrm{m}_{\textrm{L}}\right\rangle ^{2}\right],\label{eq:6}
\end{equation}

\begin{equation}
\textrm{U}_{\textrm{L}}=1-\frac{\left\langle \textrm{m}_{\textrm{L}}^{4}\right\rangle }{3\left\langle \textrm{m}_{\textrm{L}}^{2}\right\rangle ^{2}},\label{eq:7}
\end{equation}
where $\left\langle \ldots\right\rangle $ representing the thermal
average over the MCS in the stationary state, and $\textrm{m}_{\textrm{L}}$
can be either $\textrm{m}_{\textrm{L}}^{\textrm{F}}$ or $\textrm{m}_{\textrm{L}}^{\textrm{AF}}$.
We define:

\begin{equation}
m_{\alpha}^{F}=\frac{1}{N}\left\langle \sum_{i=1}^{N}S_{i}^{\alpha}\right\rangle ,\label{eq:8}
\end{equation}

\begin{equation}
m_{\alpha}^{AF}=\frac{1}{N}\left\langle \sum_{i=1}^{N}(-1)^{a+b+c}S_{i}^{\alpha}\right\rangle ,\label{eq:9}
\end{equation}
where $\alpha$ can be $x$, $y$ or $z$, and $a+b+c$ is the sum
of the spatial components of the position of $S_{i}^{\alpha}$ on
the simple cubic lattice.

Near the critical point, the Eqs. (\ref{eq:4}), (\ref{eq:5}), (\ref{eq:6})
and (\ref{eq:7}) obey the following finite-size scaling relations
\citep{21}:

\begin{equation}
\textrm{m}_{\textrm{L}}=L^{-\beta/\nu}m_{0}(L^{1/\nu}\epsilon),\label{eq:10}
\end{equation}

\begin{equation}
\textrm{\ensuremath{\chi}}_{\textrm{L}}=L^{\gamma/\nu}\chi_{0}(L^{1/\nu}\epsilon),\label{eq:11}
\end{equation}

\begin{equation}
\textrm{U}_{\textrm{L}}^{\prime}=L^{1/\nu}\frac{U_{0}^{\prime}(L^{1/\nu}\epsilon)}{\Gamma_{c}},\label{eq:12}
\end{equation}
where $\epsilon=(\Gamma-\Gamma_{c})/\Gamma_{c}$, and $\Gamma$ can
be $T$ or $q$. Thus, $\beta$, $\gamma$, and $\nu$ are the critical
exponents related the magnetization, susceptibility and correlation
length, respectively. The functions $m_{0}(L^{1/\nu}\epsilon)$, $\chi_{0}(L^{1/\nu}\epsilon)$
and $U_{0}(L^{1/\nu}\epsilon)$ are the scaling functions.
\begin{center}
\begin{figure}
\begin{centering}
\includegraphics[scale=0.7]{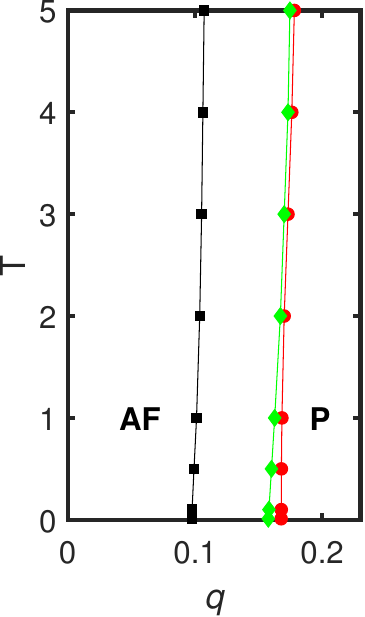}\hspace{0.1cm}\includegraphics[scale=0.7]{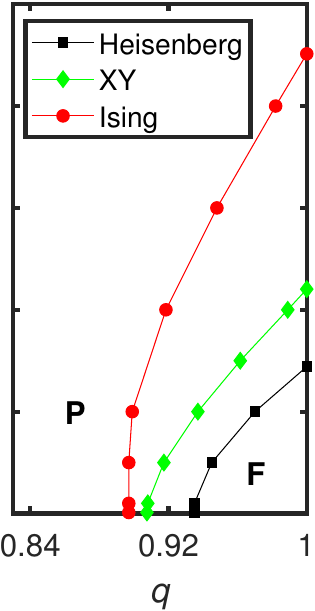}
\par\end{centering}
\caption{{\footnotesize Phase diagram of the Ising, $XY$, and Heisenberg models,
of temperature $T$ versus competition parameter $q$ on a simple
cubic lattice. There are three phases: antiferromanetic $AF$, ferromagnetic
$F$ and paramagnetic $P$. The solid lines serve only as visual guides
for the second-order phase transition points. The statistical errors
are smaller than the size of the symbols. \protect\label{fig:1}}}
\end{figure}
\par\end{center}

\section{Numerical Results and Discussion \protect\label{sec:Results}}
\begin{center}
\begin{figure*}
\begin{centering}
\includegraphics[scale=0.6]{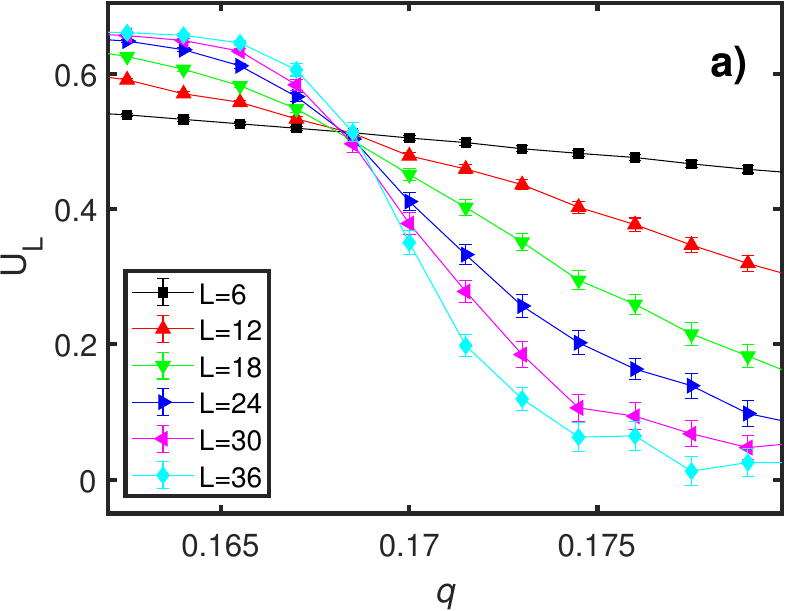}\hspace{0.25cm}\includegraphics[scale=0.6]{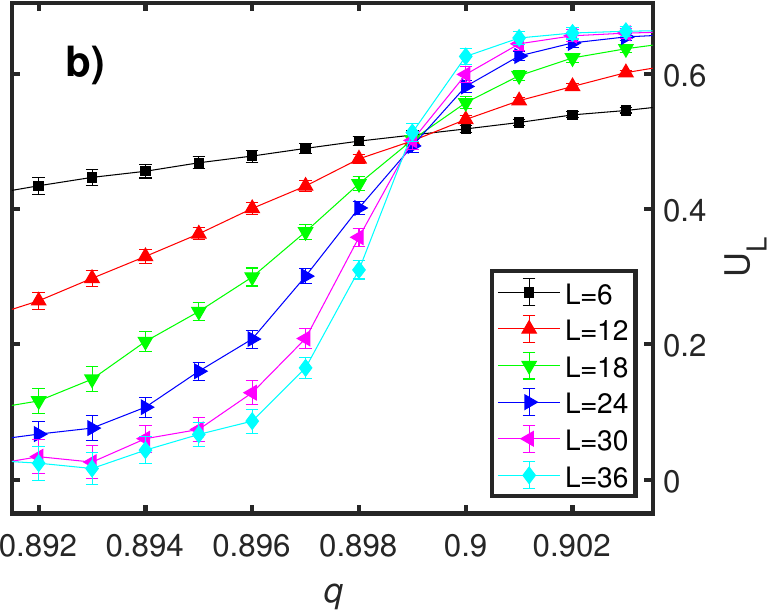}
\par\end{centering}
\begin{centering}
\includegraphics[scale=0.6]{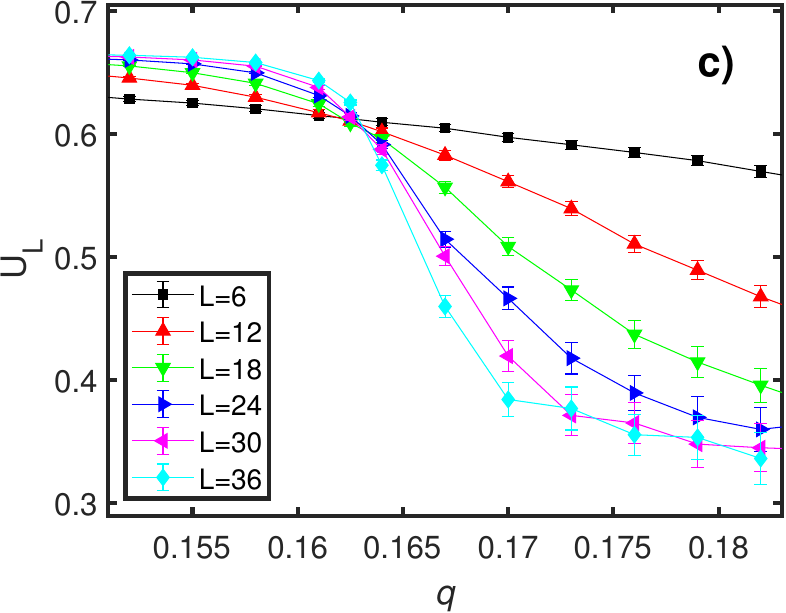}\hspace{0.25cm}\includegraphics[scale=0.6]{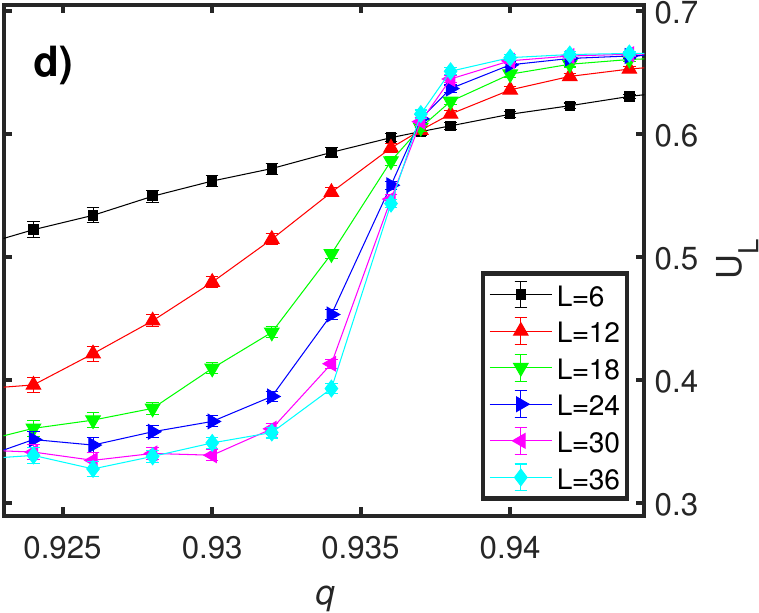}
\par\end{centering}
\begin{centering}
\includegraphics[scale=0.6]{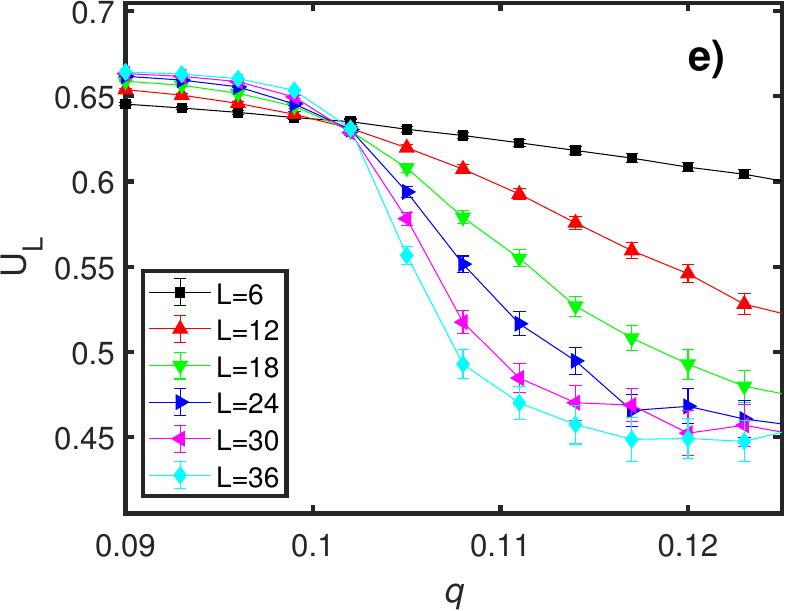}\hspace{0.25cm}\includegraphics[scale=0.6]{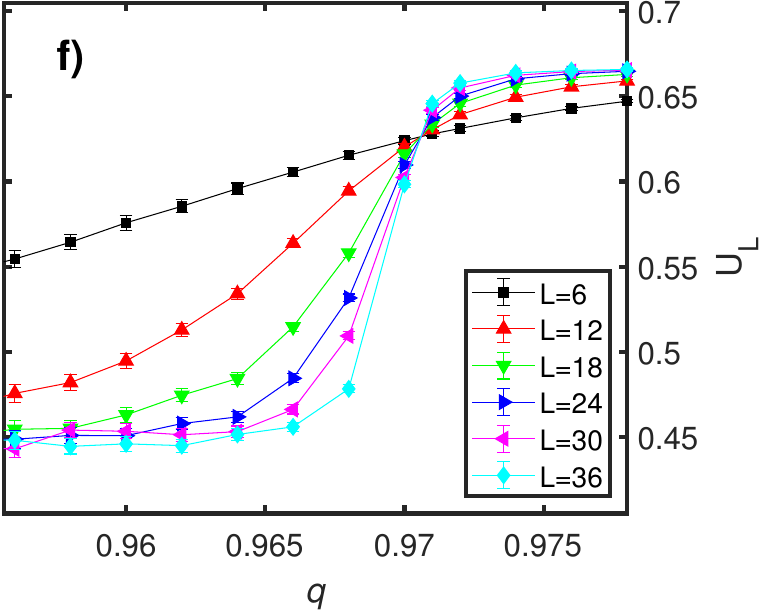}
\par\end{centering}
\caption{{\footnotesize On the left, Binder cumulant curves at the transition
from the $AF$ to the $P$ phases, while on the right, these curves
correspond to the transition from the $P$ to the $F$ phases, for
different lattice sizes $L$, as displayed in the figures. The (a)
and (b) panels are results obtained for the Ising model, (c) and (d)
for the $XY$ model, (e) and (f) for the Heisenberg model. Here, the
temperature is fixed for $T=1.0$. \protect\label{fig:2}}}
\end{figure*}
\par\end{center}

For low values of $q$, there is a high external energy flux into
the system, which drives it to the highest energy state, the antiferromagnetic
($AF$) phase. However, as this energy flux decreases, neither of
the dynamics is favored, leading the system to a disordered state
characterized by the paramagnetic ($P$) phase. At high values of
$q$, the dynamics simulating the system in contact with a thermal
reservoir dominate, and magnetic ordering is once again observed at
low temperatures, resulting in the ferromagnetic ($F$) phase, characteristic
of the lowest energy state.

In this context, we constructed the phase diagram of the system, temperature
$T$ as a function of competition parameter $q$, where the three
phases, $AF$, $P$, and $F$, are identified, as shown in Fig. \ref{fig:1}.
This figure presents the phase diagram for the models studied in the
present work: Ising, $XY$, and Heisenberg.

We observe that when the two-spin flip dynamics dominate the system,
i.e., at low $q$ values, the system reaches its highest energy state,
the $AF$ phase. However, as $q$ increases, a second-order phase
transition leads to the disordered state, the $P$ phase. Further
increasing $q$ causes the strong dominance of the one-spin flip dynamics,
which continuously organizes the system into its lowest energy state,
the $F$ phase. Unlike the $AF$ phase, the $F$ phase is strongly
temperature-dependent and occurs only at low $T$ values.

In the phase diagram shown in Fig. \ref{fig:1}, we observed that
the topology of the ordered phases remains practically unchanged regardless
of the model being considered. The only difference between the ordered
phases lies in the regions they occupy. As the spin degrees of freedom
increase in the model, they become more influenced by the thermal
fluctuations, which facilitates magnetic disordering. Consequently,
the lowest temperature or less competition between the dynamics is
required to transition into the disordered phase when dealing with
models featuring higher spin degrees of freedom.

In addition to the characteristic topology of systems subjected to
competing one- and two-spin flip dynamics, we also observed the phenomenon
of self-organization in the studied models here. This self-organization
manifests through the transitions between the ordered phases in the
system. At low values of $q$, the system resides in the $AF$ phase
and when $q$ increases, we observe a phase transition to the $P$
phase. On the other hand, at high values of $q$, we observe a phase
transition again to an ordered phase, the $F$ phase. This behavior
indicates that, beyond being independent of the lattice structure,
as shown in previous studies \citep{17,18}, the magnetic model subjected
to these reactive dynamics retains its phase transition characteristics
unchanged.

One way to identify the phase transition in the system is through
the Binder cumulant curves \citep{21,22}. This method works because,
at the critical point, the cumulant is independent of the system size.
Thus, for curves corresponding to different system sizes $L$, the
point where they intersect is identified as the phase transition point.
Moreover, if the curves exhibit continuous behavior and only take
positive values, they indicate a second-order phase transition. In
this context, we present the Binder cumulant curves associated with
the order parameter in the system for the three studied models and
for several lattice sizes $L$, can be seen in Fig. \ref{fig:2}.

In Fig. \ref{fig:2}(a), we exhibit the Binder cumulant curves for
the Ising model during the transition from the $AF$ to the $P$ phase
and in Fig. \ref{fig:2}(b) we have the curves for the same model
corresponding to the transition from the $P$ to the $F$ phase. Similarly,
in Figs. \ref{fig:2}(c) and (d), the cumulant curves for the $XY$
model are shown for the transitions from the $AF$ to the $P$ phase
and from the $P$ to the $F$ phase, respectively. Finally, in Figs.
\ref{fig:2}(e) and (f) are displayed the cumulant curves for the
Heisenberg model, also corresponding to the $AF-P$ and $P-F$ phase
transitions, respectively. In all these figures, we observe the characteristic
behavior of second-order phase transitions, along with the critical
points, $q_{c}$, identified by the intersection of the curves at
a fixed temperature of $T=1.0$.

With well-defined critical points, we can more easily calculate the
critical exponents of the system using the scaling relations from
Eqs. (\ref{eq:10}), (\ref{eq:11}), and (\ref{eq:12}), which are
valid near the criticality. To estimate the exponents, we take the
values of the quantities near the critical point as a function of
the linear lattice size $L$, plotted on a graph with logarithmic
axes. The resulting curve exhibits a linear behavior, where the slope
gives us the ratios of the critical exponents. From the $\textrm{m}_{\textrm{L}}$
curves, based in Eq. (\ref{eq:10}), we obtain the ratio $-\beta/\nu$;
from the $\textrm{\ensuremath{\chi}}_{\textrm{L}}$ curves, based
in Eq. (\ref{eq:11}), the slope returns $\gamma/\nu$. From the derivative
of the Binder cumulant near the critical point, $\textrm{U}_{\textrm{L}}^{\prime}$,
we estimate the correlation length exponent $1/\nu$, based in Eq.
(\ref{eq:12}). The linear fit of these thermodynamic quantities and
the critical exponent estimates are shown in Fig. \ref{fig:3}. In
Fig. \ref{fig:3}(a), we present the curves for the Ising model and
in Figs. \ref{fig:3}(b) and \ref{fig:3}(c) we display the curves
for the $XY$ and Heisenberg models, respectively.
\begin{center}
\begin{figure}[H]
\begin{centering}
\includegraphics[scale=0.605]{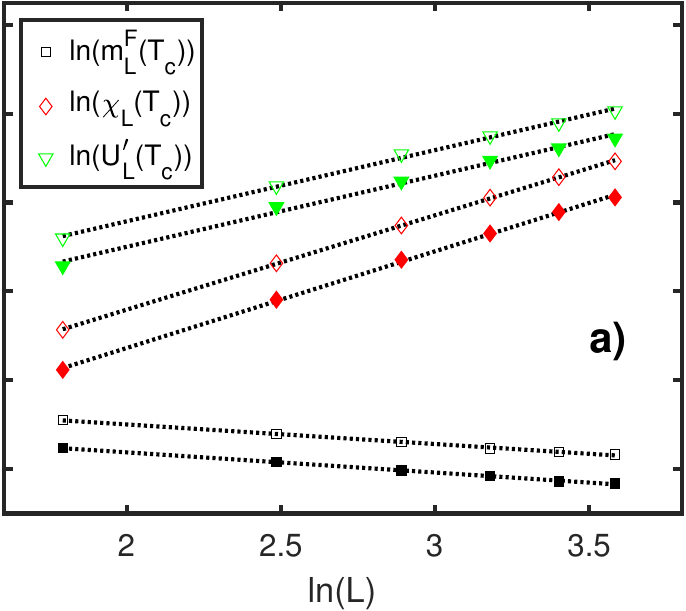}\hspace{0.32cm}
\par\end{centering}
\begin{centering}
\vspace{0.05cm}
\par\end{centering}
\begin{centering}
\includegraphics[scale=0.605]{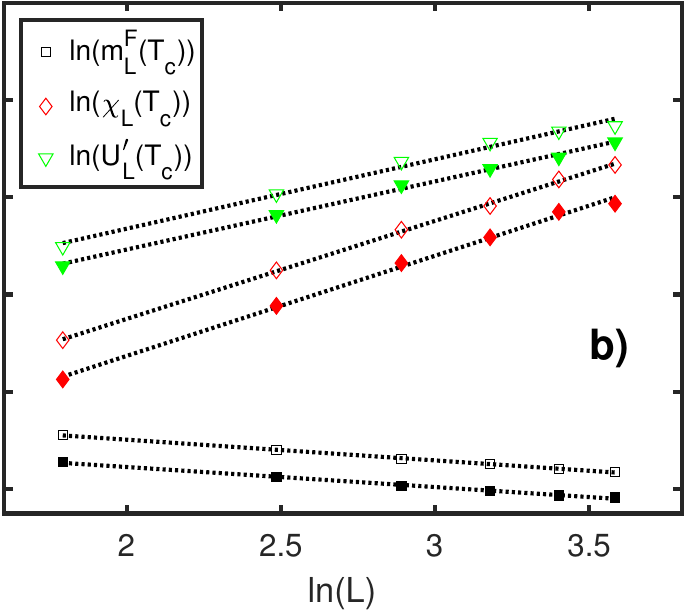}\hspace{0.32cm}
\par\end{centering}
\begin{centering}
\vspace{0.10cm}
\par\end{centering}
\begin{centering}
\includegraphics[scale=0.6]{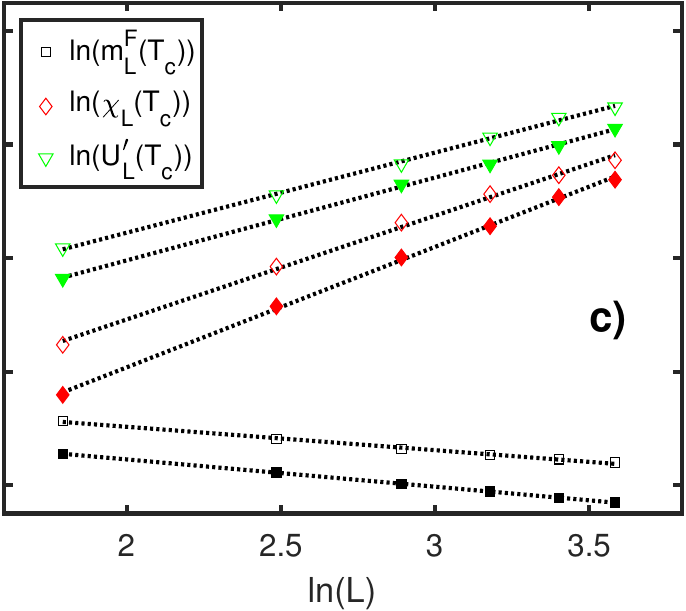}
\par\end{centering}
\caption{{\footnotesize Linear fitting of the thermodynamic quantities near
the critical point as a function of the linear lattice size $L$,
in a graph with logarithmic scales on both axes. In (a), these curves
correspond to the Ising model, in (b) the curves are obtained for
the $XY$ model, and in (c) for the Heisenberg model. \protect\label{fig:3}.}}
\end{figure}
\par\end{center}

To support these estimates, we can use data collapse for the $\textrm{m}_{\textrm{L}}$
and $\textrm{\ensuremath{\chi}}_{\textrm{L}}$curves. With data collapse,
we aim to obtain the forms of the scaling functions $m_{0}$ and $\chi_{0}$
using curves of different lattice sizes $L$. This is possible because
near the critical point if we use the correct critical point and critical
exponents in the scaling functions from Eqs. (\ref{eq:10}) and (\ref{eq:11}),
the result is a single curve, independent of the system size. To do
this, we isolate the scaling function by plotting $\textrm{m}_{\textrm{L}}L^{\beta/\nu}$
and $\textrm{\ensuremath{\chi}}_{\textrm{L}}L^{-\gamma/\nu}$ as a
function of $L^{1/\nu}\epsilon$, as shown in Fig. \ref{fig:4}. In
this case, we verify the validity of the obtained critical exponents,
as the curves with different system sizes collapse into a single curve,
representing the scaling functions. In Fig. \ref{fig:4}(a), we present
the collapsed curves for the transition from the $AF$ to the $P$
phase in the Ising model and in Fig. \ref{fig:4}(b) we have shown
the collapsed curves for the transition from the $P$ to the $F$
phase in the same model. Meanwhile, Fig. \ref{fig:4}(c) displays
the curve collapse for the $XY$ model during the $AF-P$ phase transition,
and Fig. \ref{fig:4}(d) for the $P-F$ phase transition. Lastly,
in Figs. \ref{fig:4}(e) and (f), we present the collapsed curves
for the Heisenberg model, corresponding to the $AF$ to $P$ and $P$
to $F$ phase transitions, respectively.

In order to qualitatively compare the critical exponents obtained
in this work with those reported in the literature, we present Fig.
\ref{fig:5}. This figure shows the exponents along with their respective
statistical errors, compared with the most precise results for the
exponents of the Ising \citep{23}, $XY$ \citep{24}, and Heisenberg
\citep{25} models on a simple cubic lattice under thermodynamic equilibrium.

From this comparison, we observe that the critical exponents obtained
in this study slightly deviate from those in the literature. However,
considering the statistical error margins inherent to the Metropolis
algorithm, the values align with the known exponents. Thus, we conclude
that the models in the non-equilibrium regime also preserve the system's
universality class, both for the $AF$ and $F$ phase transitions.
\begin{center}
\begin{figure*}
\begin{centering}
\includegraphics[scale=0.6]{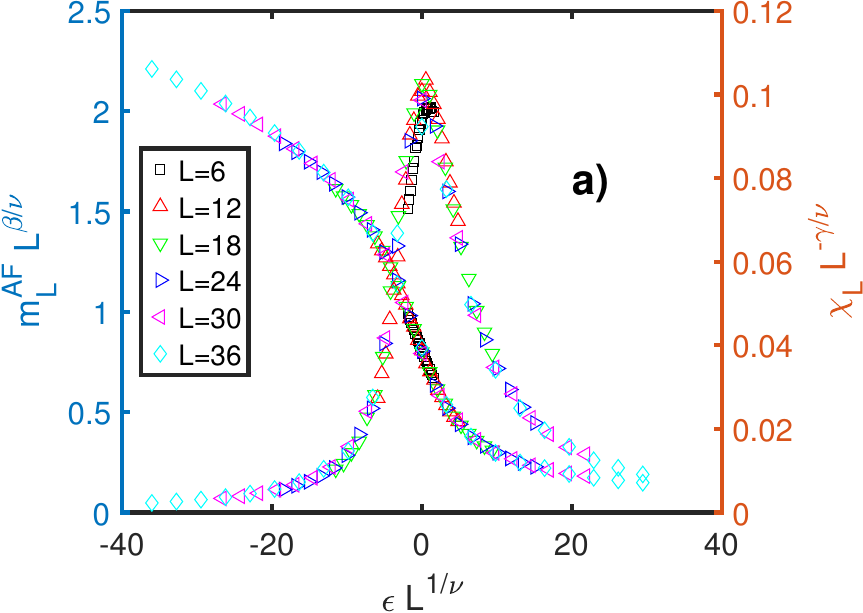}\hspace{0.25cm}\includegraphics[scale=0.6]{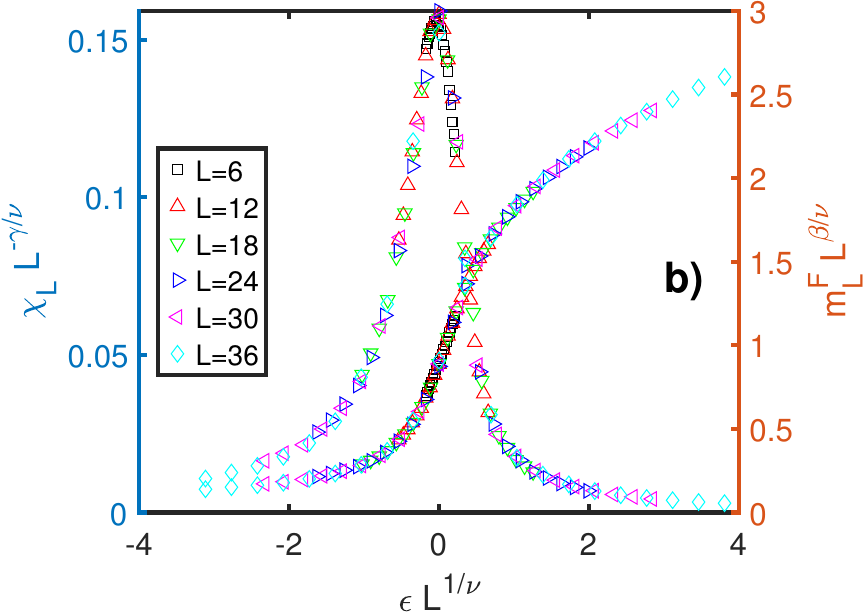}
\par\end{centering}
\begin{centering}
\includegraphics[scale=0.6]{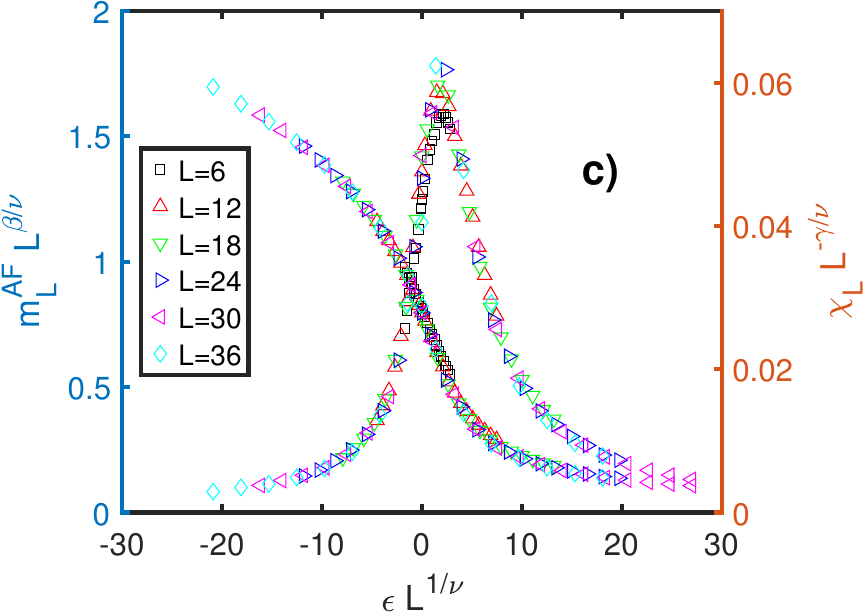}\hspace{0.25cm}\includegraphics[scale=0.6]{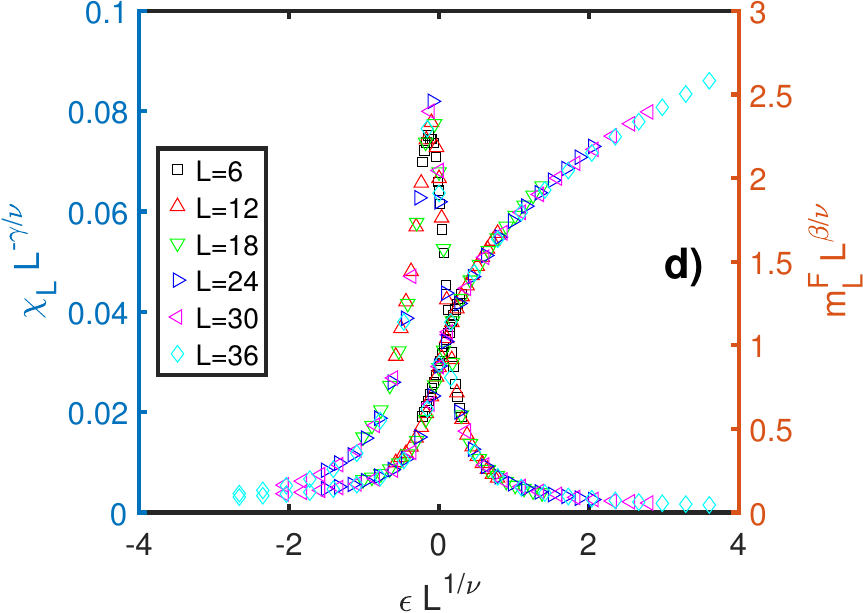}
\par\end{centering}
\begin{centering}
\includegraphics[scale=0.6]{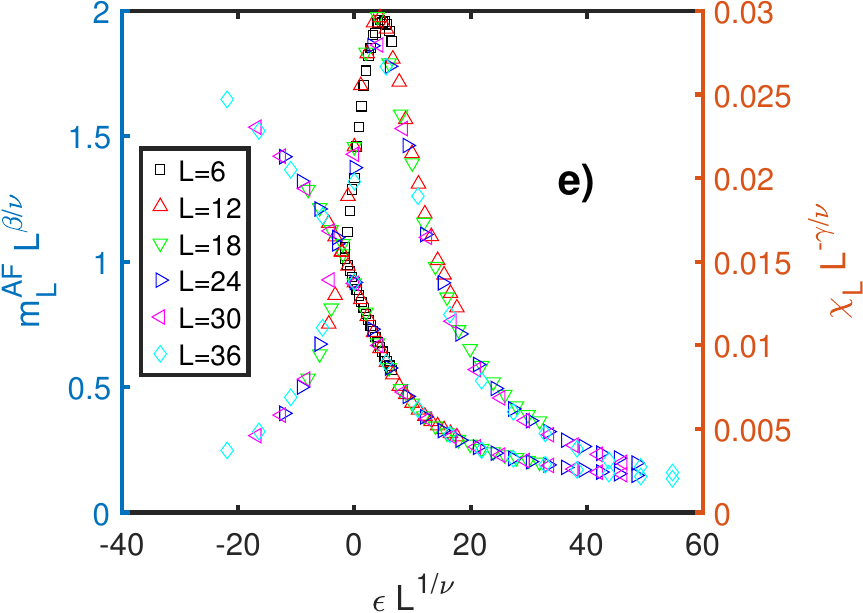}\hspace{0.25cm}\includegraphics[scale=0.6]{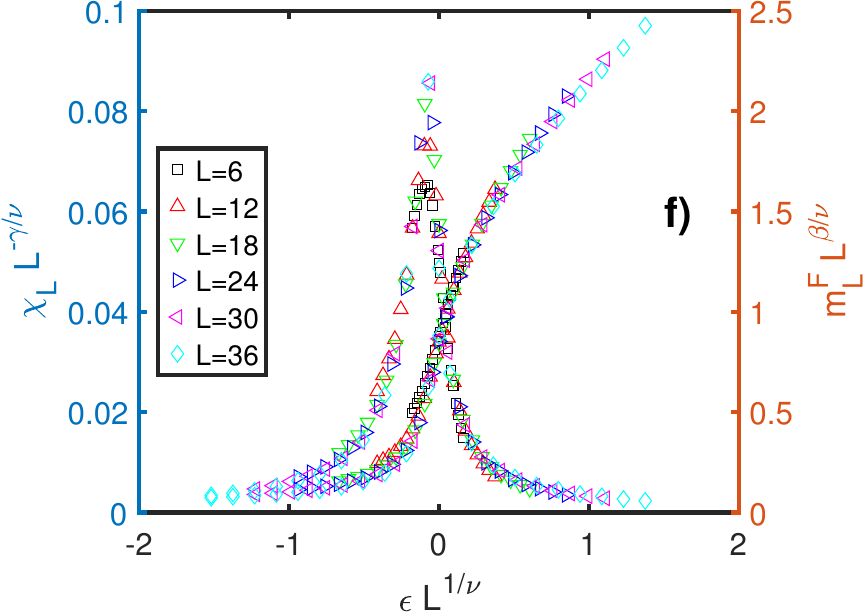}
\par\end{centering}
\caption{{\footnotesize On the left, we have the curves corresponding to the
data collapse of magnetization and magnetic susceptibility at the
transition from the $AF$ to the $P$ phases, while on the right,
these curves correspond to the transition from the $P$ to the $F$
phases, for different lattice sizes $L$, as displayed in the figures.
The (a) and (b) panels are the data collapse for the Ising model,
(c) and (d) for the $XY$ model, (e) and (f) for the Heisenberg model.
Here, the temperature is fixed for $T=1.0$. \protect\label{fig:4}}}
\end{figure*}
\par\end{center}

\begin{center}
\begin{figure}
\begin{centering}
\includegraphics[scale=0.6]{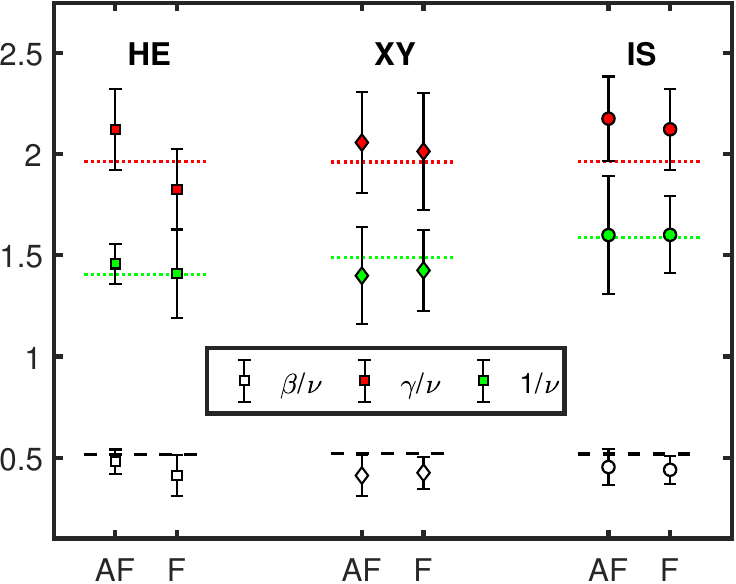}
\par\end{centering}
\caption{{\footnotesize Relation of the critical exponents obtained for the
Ising (IS), $XY$, and Heisenberg (HE) models, at the transition from
the $AF$ to the $P$ phases, and from the $P$ to the $F$ phases,
as shown on the horizontal axis. The horizontal lines correspond to
known results from the literature for these exponents \citep{23,24,25}.
\protect\label{fig:5}}}
\end{figure}
\par\end{center}

\section{Conclusions\protect\label{sec:Conclusions}}

In this work, we studied the Ising, $XY$, and Heisenberg models on
a simple cubic lattice evolving toward a stationary state through
the competition between one- and two-spin flip dynamics. With probability
$q$ , the system relaxes to its lowest energy state, as it simulates
contact with a heat bath at temperature $T$ via one-spin flip dynamics.
Conversely, with probability $1-q$, the system evolves toward a higher
energy state due to an external energy flux, governed by two-spin
flip dynamics. We constructed the phase diagram of the system, plotting
$T$ as a function of $q$ , which revealed three phases: antiferromagnetic
$AF$, paramagnetic $P$, and ferrromagnetic $F$. At low values of
$q$, where a high external energy flux is introduced into the system
due to two-spin flip dynamics, we observed the system in its highest
energy state, the $AF$ phase. As $q$ increases, the system continuously
transitions to the disordered $P$ phase. Further increasing $q$
leads to a regime where one-spin flip dynamics dominate the system,
allowing a return to an ordered phase, the $F$ phase, at low values
of $T$. This phase corresponds to the lowest energy state. We found
that the topology of the phase diagram remains unchanged regardless
of the magnetic model studied, indicating that it is determined by
the reactive dynamics used. The differences in the diagram between
models lie only in the size of the ordered phases: models with higher
spin-state degrees of freedom exhibit smaller ordered phase regions
in the diagram. Beyond phase transitions, we also calculated the system's
critical exponents and compared them with established results in the
literature. This comparison confirmed that the non-equilibrium regime
did not alter the universality class of the studied models.

\begin{acknowledgments}
This work has been supported by the Conselho Nacional de Desenvolvimento
Científico e Tecnológico (CNPq), Brazil (Process No. 140141/2024-3).
\end{acknowledgments}

\end{document}